\documentclass[seceq]{ptptex}

\usepackage{graphicx}

\title{
 Exact Results for the Jarzynski Equality
 in Ising Spin Glass Models Derived by Using a Gauge Symmetry%
}

\author{
Chiaki \textsc{Yamaguchi}%
}

\inst{
Kosugichou 1-359, Kawasaki 211-0063, Japan
}

\abst{
 Exact results for the Jarzynski equality are derived
 for Ising spin glass models.
 The Jarzynski equality is an equality
 that connects the work in nonequilibrium
 and the difference between free energies.
 The work is performed in switching an external parameter of the system.
 As the Ising spin glass models, the $\pm J$ model and 
 the Gaussian model are investigated.
 For the  $\pm J$ model and the Gaussian model,
 we derive exact lower bounds of the exponentiated work
 for investigating the ferromagnetic phases and the multicritical points,
 and derive rigorous relations between the exponentiated work
 which have different quenched random configurations.
 For the Gaussian model,
 we derive the exact exponentiated work
 for investigating the spin glass phase.
 Exact results for the infinite-range models are also obtained.
 The present results are obtained by using a gauge symmetry,
 and are related to points on the Nishimori lines
 which are special lines in the phase diagrams.
 The present results do not depend on any lattice shape, 
 and a part of the present results instead depends on
 the number of nearest-neighbor pairs in the whole system.
}

\begin{document}
\maketitle

\section{Introduction}

The theoretical studies of spin glasses have been 
 widely done \cite{KR, MPV}.
 The spin glass models have the randomness and the frustration.
 The combination of the randomness
 and the effect of frustration causes various
 interesting dynamics as well as the static properties.
 For dynamical features, the aging phenomenon,
 the dynamical transition for the distance between two spin configurations, and
 the problem of slow relaxation are reported for example \cite{KR, MPV, DW}.
 
 The Jarzynski equality is an equality that connects the work
 in nonequilibrium and the difference between free energies \cite{J1, J2}.
 The work is performed in switching an external parameter of the system.
 The Jarzynski equality is also derived in the Markov process with discrete time
 in Ref.~\citen{C1}, and
 it is pointed out in Ref.~\citen{C1} that
 the Metropolis Monte Carlo method\cite{MRRTT}
 based on the Markov process with discrete time
 is a suited example for applying the Jarzynski equality.
 Also, the Metropolis Monte Carlo method and
 the related Monte Carlo methods are known as powerful methods to
 investigate the spin glass models on finite-dimensional lattices\cite{KR}.
 Therefore, the application of the Jarzynski equality and the related theories
 to the spin glass models may make its dynamical features clearer.
 Several exact results for applying the Jarzynski equality to the $\pm$ J Ising spin glass model
 are also shown in Ref.~\citen{ON}, and
 it is suggested in Ref.~\citen{ON} that
 the application of the Jarzynski equality and the related theories
 to the spin glass models may be useful for avoiding the problem of slow relaxation,
 that reaching the equilibrium states is hard, in spin glass models.

As the Ising spin glass models, the $\pm J$ model \cite{N, KR} and
 the Gaussian model \cite{N, MPV, SK} are investigated.
 These models
 have different quenched randomness.
 In Ref.~\citen{KR}, for the $\pm J$ model, a number of studies by computer simulations
 are seen.
 In Ref.~\citen{MPV}, for the Gaussian model, a number of analytical studies
 are seen.
 The present results for the infinite-range models are also shown.

In this article,
 a gauge transformation is used.
 There are special lines
 in the phase diagrams for several spin glass models,
 where the lines are called the Nishimori lines \cite{N}.
 Several physical quantities and several bounds
 for physical quantities
 are exactly calculated on the Nishimori lines
 by using the gauge transformation \cite{N}.
 The present results are related to points on the Nishimori lines.
 In the applications of many other methods,
 a lattice shape is supposed in advance, and
 the results are calculated on the lattice.
 On the other hand, in the method applied in this article,
 any lattice shape is not supposed in advance.
 The present results do not depend on any lattice shape, and
 a part of the present results instead depends on
 the number of nearest-neighbor pairs in the whole system.
 The present results are exact.

 In this article, we exactly analytically investigate
 the exponentiated work in the Jarzynski equality.
 The present results are useful
 for understanding the exact exponentiated work
 in the phase diagrams for the Ising spin glass models.
 In addition, the present results are
 useful for estimating the results by
 approximation approaches, and are useful for fixing bugs
 in computer programs for investigating the Ising spin glass models.

This article is organized as follows.
 The Jarzynski equality and the models are explained in \S\ref{sec:2}.
 Exact results for the $\pm J$ model are given in \S\ref{sec:3}, and
 exact results for the Gaussian model are given in \S\ref{sec:4}.
 This article is summarized in \S\ref{sec:5}.

\section{The Jarzynski equality and the models} \label{sec:2}

We explain the Jarzynski equality \cite{J1, J2, C1}.
 We define the inverse temperature of the reservoir as $\beta$.
 Here, $\beta = 1 / k_B T$, $T$ is the temperature,
 and $k_B$ is the Boltzmann constant.
 We define a parameter for
 the strength of the exchange interaction between spins as $J ( > 0)$. 
 We use a representation: $K = \beta J$.
 By using $K_a$ for $K = K_a$ and $K_b$ for $K = K_b$,
 we consider a nonequilibrium process from $K_a$ to $K_b$.
 We assume that the initial and final states are in equilibrium, 
 and the states in the process from $K_a$ to $K_b$ are in nonequilibrium.
 The Jarzynski equality is given by \cite{J1, J2, C1} 
\begin{equation}
\overline{e^{- \beta W}} = e^{ - \beta \Delta F} \, ,
 \label{eq:Jeft0}
\end{equation}
 where 
 $W$ is the work performed in the process from $K_a$ to $K_b$, and
 the overbar indicates an ensemble average over all
possible paths through phase space. $\Delta F$ is 
 the difference between free energies given by $\Delta F = F (K_b ) - F (K_a )$,
 where $F (K_a ) $ is the free energy for $K = K_a$.
 The left-hand side of Eq.~(\ref{eq:Jeft0}) is the 
 nonequilibrium measurements, and
 the right-hand side of Eq.~(\ref{eq:Jeft0}) is the 
 equilibrium information.
 By using Eq.~(\ref{eq:Jeft0}),
 one can extract the equilibrium information from
 the ensemble of nonequilibrium.
 Eq.~(\ref{eq:Jeft0}) does not depend on
 both the path from $K_a$ to $K_b$, and the rate at which
 the parameters are switched along the path.
 Note that, here, $W$ is the work for changing $J$ and a fixed $\beta$.
 The work in switching an external parameter of the system
 in contact with a heat reservoir is originally supposed\cite{J1, J2}.
 In Ref.~\citen{J1}, some particle-particle interactions which are turned on or off
during the course of a molecular dynamics simulation are
mentioned as a suited example for the applications of the Jarzynski equality,
and the present study may be exactly included in the example.
 As for the dynamics of the nonequilibrium process,
 the Markov process with discrete time is considered in this study, for example\cite{C1}. 
 Eq.~(\ref{eq:Jeft0}) is equivalently given by
\begin{equation}
\overline{e^{- \beta W}} = \frac{Z (K_b )}{Z (K_a ) } \, ,
 \label{eq:Jeft1}
\end{equation}
 where $Z (K_a )$ is the partition function for $K = K_a$.

The Hamiltonian for Ising spin glass models, ${\cal H}$, 
 is given by  \cite{KR, N, MPV, SK}
\begin{equation}
 {\cal H} = - J \sum_{\langle i, j \rangle} \tau_{i, j} S_i S_j \, ,
\end{equation}
 where $\langle i, j \rangle$ denotes nearest-neighbor pairs, $S_i$ is
 a state of the spin at the site $i$, and $S_i = \pm 1$.
 $J \tau_{i, j}$ is the strength of the exchange interaction
 between the spins at the sites $i$ and $j$. Here, $J > 0$,
 and the value of $J$ is switched from $K_a / \beta$ to $K_b / \beta$.
 The value of $\tau_{i, j}$ is given with a distribution $P ( \tau_{i, j})$.
 The $\pm J$ model and the Gaussian model
 are given by different distributions of $\tau_{i, j}$.

 The distribution $P^{(\pm J)} ( \tau_{i, j})$ of $\tau_{i, j}$
 for the $\pm J$ model is given by \cite{N, KR}
\begin{equation}
 P^{(\pm J)} ( \tau_{i, j})
 = p \, \delta_{ \tau_{i, j}, 1} + (1 - p) \, \delta_{ \tau_{i, j}, - 1} \, ,
 \label{eq:PpmJJij}
\end{equation}
 where $\delta$ is the Kronecker delta.
 $p$ is the probability that the interaction
 is ferromagnetic, and $1 - p$ is
 the probability that the interaction is antiferromagnetic. 
 By using Eq.~(\ref{eq:PpmJJij}), the distribution $P^{(\pm J)} ( \tau_{i, j})$ 
 is rewritten as \cite{N}
\begin{equation}
 P^{(\pm J)} ( \tau_{i, j}) = \frac{e^{K^{(\pm J)}_p \tau_{i, j}} }
{2 \cosh K^{(\pm J)}_p} \, , \quad \tau_{i, j} = \pm 1 \label{eq:Ptauij} \, ,
\end{equation}
 where $K^{(\pm J)}_p$ is given by
\begin{equation}
 K^{(\pm J)}_p = \frac{1}{2} \ln \frac{p}{1-p} \, . \label{eq:betaPpmJ}
\end{equation}
 When the value of $K^{(\pm J)}_p$ is consistent with the value of  $K$,
 the line for $K = K^{(\pm J)}_p$ in the phase diagram
 is called the Nishimori line.
 The distribution $P^{({\rm G})} ( \tau_{i, j})$ of $\tau_{i, j}$
 for the Gaussian model is given by \cite{N, MPV, SK}
\begin{equation}
 P^{({\rm G})} ( \tau_{i, j})
 = \frac{1}{\sqrt{2 \pi J^2_1}} \, \exp \biggl[ - \frac{(\tau_{i, j} - J_0 )^2 }
 {2 J^2_1 } \biggr]  \, ,
 \label{eq:PGaussianJij}
\end{equation}
 and is rewritten as \cite{N}
\begin{equation}
 P^{({\rm G})} ( \tau_{i, j})
 = \frac{1}{\sqrt{2 \pi J^2_1}} \,
 \exp \biggl\{ -  \frac{ \tau^2_{i, j} }{ 2 J^2_1 }
  - \frac{ [ K^{({\rm G}) }_p J_1 ]^2 }{ 2}
 + K^{({\rm G})}_p \tau_{i, j} \biggr\}  \, ,
 \label{eq:PGaussianJij2}
\end{equation}
 where $K^{({\rm G})}_p$ is given by
\begin{equation}
 K^{({\rm G})}_p =  \frac{J_0 }{J^2_1 } \, . \label{eq:betaPGaussianJ}
\end{equation}
 When the value of $K^{({\rm G})}_p$ is consistent with the value of  $K$, 
 the line for $K = K^{({\rm G })}_p$ in the phase diagram
 is also called the Nishimori line.

A gauge transformation \cite{N, T} given by
\begin{equation}
 J \tau_{i, j} \to J \tau_{i, j} \sigma_i \sigma_j \, , \quad S_i \to S_i \sigma_i 
 \label{eq:GaugeT} 
\end{equation}
 is used, where $\sigma_i = \pm 1$.
 The gauge transformation has no effect on thermodynamic quantities\cite{T}.
 By using the gauge transformation,
 the Hamiltonian  ${\cal H}$ part becomes ${\cal H} \to {\cal H}$, and
 the distribution $P^{(\pm J)} (\tau_{i, j})$ part becomes
\begin{equation}
 \prod_{\langle i, j \rangle} P^{(\pm J)} (\tau_{i, j})
 \to \frac{\sum_{\{ \sigma_i \}}
 e^{K^{(\pm J)}_p \sum_{\langle i, j \rangle }
 \tau_{i, j} \sigma_i \sigma_j  } }
 {2^ N (2 \cosh K^{(\pm J)}_p)^{N_B}} 
 = \frac{Z (K^{(\pm J)}_p )  }
 {2^ N (2 \cosh K^{(\pm J)}_p)^{N_B}} \, ,
 \label{eq:Ptauij2pmJ}
\end{equation}
 where $N$ is the number of sites, and $N_B$ is the number of
 nearest-neighbor pairs in the whole system.
 By using the gauge transformation,
 the distribution $P^{({\rm Gaussian})} (\tau_{i, j})$ part becomes
\begin{eqnarray}
 & & \prod_{\langle i, j \rangle} P^{({\rm G})} (\tau_{i, j})
 \nonumber \\
 &\to& \frac{ 1}{2^N (2 \pi J^2_1)^{\frac{N_B }{2 }} } \,
 \sum_{ \{ \sigma_i \} } \exp \biggl\{ - \sum_{\langle i, j \rangle }
  \frac{ \tau^2_{i, j} }{ 2 J^2_1 }
  -  \frac{ N_B [ K^{({\rm G}) }_p J_1 ]^2 }{2}
 +  K^{({\rm G})}_p
 \sum_{\langle i, j \rangle } \tau_{i, j} \sigma_i \sigma_j \biggr\} 
 \nonumber \\
 &=& \frac{ 1}{2^N (2 \pi J^2_1)^{\frac{N_B }{2 } } } \,
 \exp \biggl\{ - \sum_{\langle i, j \rangle }
 \frac{ \tau^2_{i, j} }{ 2 J^2_1 }
  - \frac{  N_B [ K^{({\rm G})}_p J_1 ]^2 }{2}
 \biggr\} Z (K^{({\rm G})}_p )
 \, .
 \label{eq:Ptauij2Gaussian}
\end{eqnarray}

\section{Exact results for the $\pm J$ Ising spin glass model} \label{sec:3}

 We derive exact results for the $\pm J$ Ising spin glass model.
 We multiply both sides of Eq.~(\ref{eq:Jeft1}) by
$$
\frac{ e^{K_a \sum_{ \langle i, j \rangle } \tau_{i, j} }}
{(2 \cosh K_a )^{N_B}} \, ,
$$ 
 and perform the summation over $\{ \tau_{i, j} \} $. Then, we obtain
\begin{equation}
 [ \overline{e^{- \beta W}} ]_{K_a}
 =   \biggl( \frac{2 \cosh K_b}{2 \cosh K_a} \biggr)^{N_B} \sum_{\{ \tau_{i, j} \} }  
 \frac{ e^{K_a \sum_{\langle i, j \rangle } \tau_{i, j} } }{ (2 \cosh K_b )^{N_B} }
 \frac{Z (K_b )}{Z (K_a ) } \, ,
 \label{eq:PJeft1}
\end{equation}
 where $[ \, ]_{K_a }$ denotes the quenched random configuration average for $K_p = K_a$.
 By  performing the gauge transformation to the right-hand side of Eq.~(\ref{eq:PJeft1}),
 we obtain \cite{ON}
\begin{equation}
 [  \overline{e^{- \beta W}} ]_{K_a}
 =  \biggl( \frac{2 \cosh K_b}{2 \cosh K_a} \biggr)^{N_B} \, .
 \label{eq:PJeft2}
\end{equation}
 This equation is exact, is derived in Ref.~\citen{ON},
 and is called the Jarzynski equality for spin glass \cite{ON}.
 When $K_a$ is fixed,
 the exponentiated work
 $[ \overline{e^{- \beta W}} ]_{K_a}$ is monotonically increasing for $K_b$.
 The exponentiated work $[ \overline{e^{- \beta W}} ]_{K_a}$
 does not depend on any lattice shape, and instead
 depends on the number of nearest-neighbor pairs in the whole system, $N_B$.
 This equation is useful for estimating the results by approximation
 approaches, and is useful for fixing bugs in computer programs
 for investigating this model.

 When $J \to J / \sqrt{N}$ and $N_B = N ( N - 1 ) /2$ are set, 
 the model becomes the infinite-range model.
 We obtain
\begin{equation}
 [  \overline{e^{- \beta W}} ]_{K_a}
 \to  \exp \biggl[ \frac{(N - 1) (K^2_b - K^2_a ) }{4} \biggr] \, 
 \label{eq:PJeft3}
\end{equation}
 for the infinite-range model with $N \to \infty$.
 This equation can be called the Jarzynski equality
 for the infinite-range $\pm J$ Ising spin glass model.
 When $K_a$ is fixed,
 the exponentiated work $[ \overline{e^{- \beta W}} ]_{K_a}$ with $N \to \infty$
 is monotonically increasing for $K_b$.

 When $K_a = 0$ and $K_b = K$ are set
 in Eqs.~(\ref{eq:PJeft2}) and (\ref{eq:PJeft3}), we obtain
 $[ \overline{e^{- \beta W}} ]_{K_p = 0} = \exp [ N_B \ln ( \cosh K  ) ]$ 
 for the finite-dimensional model and
 $[  \overline{e^{- \beta W}} ]_{K_p = 0} \to \exp [ (N - 1)  K^2 / 4 ] $
 for the infinite-range model with $N \to \infty$.
 These result are results for $p = 1 / 2$, thus
 the randomness and the effect of frustration are largest in this model.
 In addition,
 these results are results for the nonequilibrium process from $K = 0$ to $K = K$.
 Therefore, these results are results under simple conditions.

\begin{figure}[t]
\begin{center}
\includegraphics[width=0.68\linewidth]{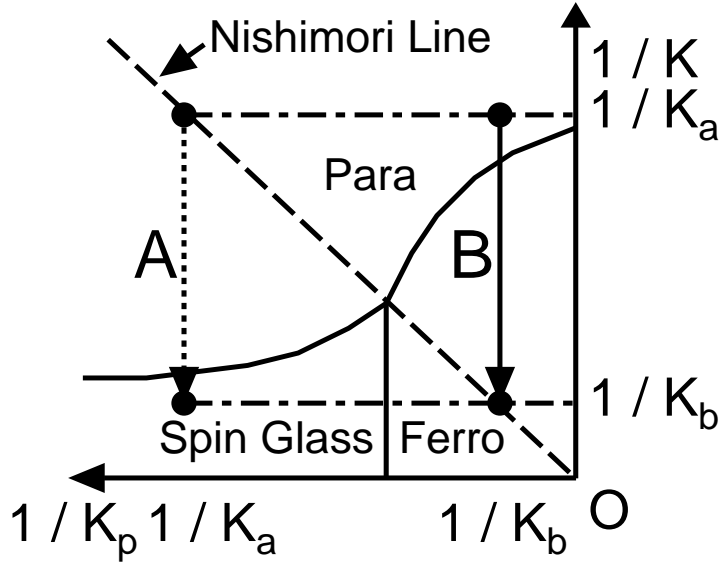}
\end{center}
\caption{
 Two nonequilibrium processes and a schematic phase diagram 
 for the $\pm J$ Ising spin glass model.
 $K_p$ is a parameter for the concentrations of the 
 ferromagnetic interaction and the antiferromagnetic interaction.
 $K$ is a parameter for the inverse temperature and 
 the strength of the exchange interaction.
 The paramagnetic phase (`Para'),
 the ferromagnetic phase (`Ferro')
 and the spin glass phase (`Spin Glass') are depicted.
 The Nishimori line is also depicted as the dashed line.
 The nonequilibrium process for the exponentiated work $[  \overline{e^{- \beta W}} ]_{K_a}$
 is depicted as the arrow `A'.
 The nonequilibrium process for the exponentiated work $[  \overline{e^{- \beta W}} ]_{K_b}$
 is depicted as the arrow `B'.
\label{fig:phase-diagram}
}
\end{figure}
 Fig.~\ref{fig:phase-diagram} shows
 two nonequilibrium processes and a schematic phase diagram 
 for the $\pm J$ model.
 $K_p$ is a parameter for the concentrations of the 
 ferromagnetic interaction and the antiferromagnetic interaction, and
 is given in Eq.~(\ref{eq:betaPpmJ}).
 $K$ is a parameter for the inverse temperature and
 the strength of the exchange interaction, and
 is given by $K = \beta J$.
 The paramagnetic phase (`Para'),
 the ferromagnetic phase (`Ferro')
 and the spin glass phase (`Spin Glass') are depicted.
 The Nishimori line is also depicted as the dashed line.
 The nonequilibrium process for the exponentiated work $[  \overline{e^{- \beta W}} ]_{K_a}$
 is depicted as the arrow `A', and is mentioned above and in Ref.~\citen{ON}.
 The nonequilibrium process for the exponentiated work $[  \overline{e^{- \beta W}} ]_{K_b}$
 is depicted as the arrow `B', and is mentioned below.

The Cauchy-Schwarz inequality is given by
\begin{equation}
  [ \sum_{ \{ \tau_{i, j} \} } x ( \{ \tau_{i, j} \}  ) y (\{ \tau_{i, j} \}  ) ]^2
 \leq  \sum_{ \{ \tau_{i, j} \} } x^2 ( \{ \tau_{i, j} \}  ) \cdot
 \sum_{ \{ \tau_{i, j} \} } y^2 (\{ \tau_{i, j} \}  ) 
 \, 
 \label{eq:Cauchy-Schwarz}
\end{equation}
 for example, where $x  ( \{ \tau_{i, j} \}  ) $ and  $y  ( \{ \tau_{i, j} \}  ) $ are variables which depend on $\{ \tau_{i, j} \}$. 
When  $x  ( \{ \tau_{i, j} \}  ) $ and  $y  ( \{ \tau_{i, j} \}  ) $ are
\begin{equation}
 x  ( \{ \tau_{i, j} \}  ) = \sqrt{
 \frac{ e^{K_a \sum_{\langle i, j \rangle } \tau_{i, j} } }{ (2 \cosh K_b )^{N_B} } } 
 \, 
 \label{eq:PJeft4}
\end{equation}
and
\begin{equation}
 y  ( \{ \tau_{i, j} \}  ) = \sqrt{
 \frac{ e^{K_a \sum_{\langle i, j \rangle } \tau_{i, j} } }{ (2 \cosh K_b )^{N_B} } }
 \frac{Z (K_b )}{Z (K_a ) } 
 \, ,
 \label{eq:PJeft5}
\end{equation}
 by applying the Cauchy-Schwarz inequality and the gauge transformation, we obtain
\begin{equation}
  \biggl( \frac{2 \cosh K_b}{2 \cosh K_a} \biggr)^{N_B}
 \leq  
  [  \overline{e^{- \beta W}} ]_{K_b}
 \, .
 \label{eq:PJeft6}
\end{equation}
 The left-hand side of Eq.~(\ref{eq:PJeft6}) is a lower bound
 of the exponentiated work $[  \overline{e^{- \beta W}} ]_{K_b}$.
 The  lower bound of the exponentiated work $[  \overline{e^{- \beta W}} ]_{K_b}$
 does not depend on any lattice shape,
 and instead depends on the number of nearest-neighbor pairs in the whole system, $N_B$.
 This relation is rigorous, is useful for estimating the results by approximation
 approaches, and is useful for fixing bugs in computer programs
 for investigating this model.

 For the infinite-range model with $N \to \infty$, we obtain
\begin{equation}
   \exp \biggl[ \frac{(N - 1) (K^2_b - K^2_a ) }{4} \biggr] 
 \leq  
  [  \overline{e^{- \beta W}} ]_{K_b}
 \, .
 \label{eq:PJeft7}
\end{equation}

The nonequilibrium process for the exponentiated work $[ \overline{e^{- \beta W}} ]_{K_a}$
 is limited to the under left side of the Nishimori line in the phase diagram
 shown as the arrow `A' in Fig.~\ref{fig:phase-diagram}.
 Therefore, Eqs.~(\ref{eq:PJeft2}) and (\ref{eq:PJeft3})
 are useful for investigating the spin glass phase.
 The nonequilibrium process for the exponentiated work $[ \overline{e^{- \beta W}} ]_{K_b}$
 is limited to the upper right side of the Nishimori line in the phase diagram
 shown as the arrow `B' in Fig.~\ref{fig:phase-diagram}.
 Therefore, the left-hand sides of Ineqs.~(\ref{eq:PJeft6}) and (\ref{eq:PJeft7}) 
 are useful as lower bounds
 in investigating the ferromagnetic phase and the multicritical point
 for the paramagnetic phase, the ferromagnetic phase and the spin glass phase.
 
By using Eqs.~(\ref{eq:PJeft2}) and (\ref{eq:PJeft6}),
 we obtain
\begin{equation}
 [  \overline{e^{- \beta W}} ]_{K_a}  \leq 
   [  \overline{e^{- \beta W}} ]_{K_b} \, .
 \label{eq:PJeft8}
\end{equation}
 The relation (\ref{eq:PJeft8})
 does not depend on any lattice shape,
 and also holds for the infinite-range model.
 This relation is a rigorous relation between the exponentiated work
 which have different quenched random configurations.
 This relation is useful for understanding the exact exponentiated work
 in the phase diagram.
 This relation exactly shows that
 the exponentiated work of the nonequilibrium process shown
 as the arrow `B' in Fig.~\ref{fig:phase-diagram} is greater than or equal to
 that of the nonequilibrium process shown
 as the arrow `A' in Fig.~\ref{fig:phase-diagram}.

\section{Exact results for the Gaussian Ising spin glass model} \label{sec:4}

 We derive exact results for the Gaussian Ising spin glass model.
 We multiply both sides of Eq.~(\ref{eq:Jeft1}) by
$$
 \frac{ 1}{(2 \pi J^2_1)^{\frac{N_B }{2 }} } \,
  \exp \biggl[ - \sum_{\langle i, j \rangle }
  \frac{ \tau^2_{i, j} }{ 2 J^2_1 }
  -  \frac{ N_B ( K_a J_1 )^2 }{2}
  +  K_a
  \sum_{\langle i, j \rangle } \tau_{i, j} \biggr] \, ,
$$
 and perform the integration over $\{ \tau_{i, j} \} $. Then, we obtain
\begin{eqnarray}
  [  \overline{e^{- \beta W}} ]_{K_a}
 &=& \frac{ 1}{(2 \pi J^2_1)^{\frac{N_B }{2 }} } \,
  e^{- \frac{ N_B ( K_a J_1 )^2 }{2}  }
 \nonumber \\ & & \times
 \int^{\infty }_{ - \infty } \cdots \int^{\infty }_{ - \infty }
 ( \prod_{\langle i, j \rangle } d \tau_{i, j} ) 
  \, e^{- \sum_{\langle i, j \rangle } \frac{ \tau^2_{i, j} }{ 2 J^2_1 }
  + K_a \sum_{\langle i, j \rangle } \tau_{i, j} }
 \frac{Z (K_b )}{Z (K_a ) } \, .
 \label{eq:GaJeft1}
\end{eqnarray}
 By  performing the gauge transformation
 to the right-hand side of Eq.~(\ref{eq:GaJeft1}), we obtain
\begin{equation}
 [  \overline{e^{- \beta W}} ]_{K_a}
 = \exp \biggl[ \frac{ N_B ( K^2_b - K^2_a ) J^2_1  }{2} \biggr] \, .
 \label{eq:GaJeft2}
\end{equation}
 This equation is exact,
 and can be called the Jarzynski equality for the Gaussian Ising spin glass model.
 When $K_a$ is fixed,
 the exponentiated work $[ \overline{e^{- \beta W}} ]_{K_a}$
 is monotonically increasing for $K_b$.
 The exponentiated work $[ \overline{e^{- \beta W}} ]_{K_a}$
 does not depend on any lattice shape, and instead
 depends on the number of nearest-neighbor pairs in the whole system, $N_B$.
 This result is useful for estimating the results by approximation
 approaches, and is useful for fixing bugs in computer programs
 for investigating this model.

 When $J \to J / \sqrt{N}$ and $N_B = N ( N - 1 ) /2$ are set,
 the model becomes the infinite-range model, i.e., the Sherrington-Kirkpatrick model \cite{SK}.
 We obtain
\begin{equation}
 [  \overline{e^{- \beta W}} ]_{K_a}
 =  \exp \biggl[ \frac{(N - 1) (K^2_b - K^2_a ) J^2_1 }{4} \biggr] \, 
 \label{eq:GaJeft3}
\end{equation}
 for the Sherrington-Kirkpatrick model.
 This equation can be called the Jarzynski equality
 for the Sherrington-Kirkpatrick model.
 When $K_a$ is fixed,
 the exponentiated work
 $[ \overline{e^{- \beta W}} ]_{K_a}$ is monotonically increasing for $K_b$.

 When $K_a = 0$ and $K_b = K$ are set
 in Eqs.~(\ref{eq:GaJeft2}) and (\ref{eq:GaJeft3}), we obtain
 $[ \overline{e^{- \beta W}} ]_{K_p = 0} = \exp [ N_B ( K J_1 )^2 / 2 ]$ 
 for the finite-dimensional model and
 $[ \overline{e^{- \beta W}} ]_{K_p = 0} = \exp [ (N - 1) ( K J_1 )^2 / 4 ]$
 for the infinite-range model.
 These result are results for $J_0 = 0$, thus
 the randomness and the effect of frustration are largest in this model.
 In addition,
 these results are results for the nonequilibrium process from $K = 0$ to $K = K$.
 Therefore, these results are results under simple conditions.

\begin{figure}[t]
\begin{center}
\includegraphics[width=0.68\linewidth]{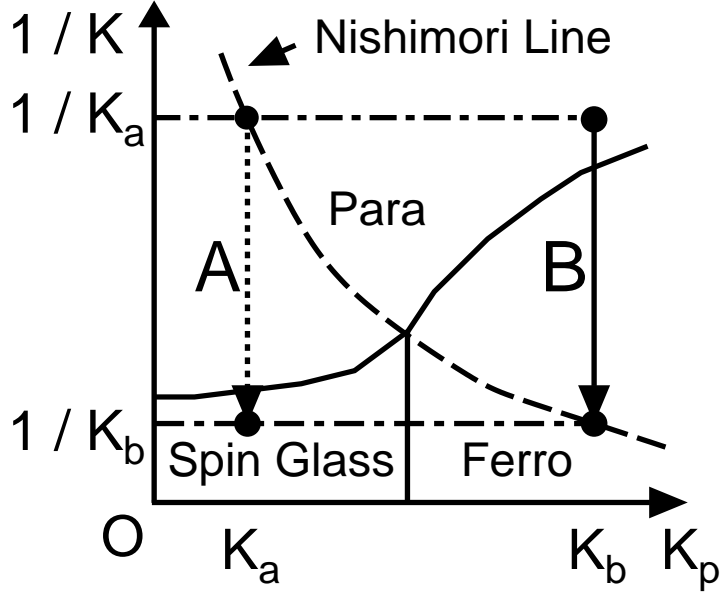}
\end{center}
\caption{
 Two nonequilibrium processes and a schematic phase diagram 
 for the Gaussian Ising spin glass model.
  $K_p$ is a parameter for the concentrations of the
 ferromagnetic interaction and the antiferromagnetic interaction.
 $K$ is a parameter for the inverse temperature and
 the strength of the exchange interaction.
 The paramagnetic phase (`Para'),
 the ferromagnetic phase (`Ferro')
 and the spin glass phase (`Spin Glass') are depicted.
 The Nishimori line is also depicted as the dashed line.
 The nonequilibrium process for the exponentiated work $[  \overline{e^{- \beta W}} ]_{K_a}$
 is depicted as the arrow `A'.
 The nonequilibrium process for the exponentiated work $[  \overline{e^{- \beta W}} ]_{K_b}$
 is depicted as the arrow `B'.
\label{fig:phase-diagramGa}
}
\end{figure}
 Fig.~\ref{fig:phase-diagramGa} shows
 two nonequilibrium processes and a schematic phase diagram 
 for the Gaussian model.
 $K_p$ is a parameter for the concentrations of the 
 ferromagnetic interaction and the antiferromagnetic interaction, and
 is given in Eq.~(\ref{eq:betaPGaussianJ}).
 $K$ is a parameter for the inverse temperature and
 the strength of the exchange interaction, and
 is given by $K = \beta J$.
 The paramagnetic phase (`Para'),
 the ferromagnetic phase (`Ferro')
 and the spin glass phase (`Spin Glass') are depicted.
 The Nishimori line is also depicted as the dashed line.
 The nonequilibrium process for the exponentiated work $[  \overline{e^{- \beta W}} ]_{K_a}$
 is depicted as the arrow `A', and is mentioned above.
 The nonequilibrium process for the exponentiated work $[  \overline{e^{- \beta W}} ]_{K_b}$
 is depicted as the arrow `B', and is mentioned below.

The Cauchy-Schwarz inequality is given by
\begin{eqnarray}
  & & [ \int^{\infty }_{ - \infty } \cdots \int^{\infty }_{ - \infty } 
 ( \prod_{\langle i, j \rangle } d \tau_{i, j} ) \,
  x ( \{ \tau_{i, j} \}  ) y (\{ \tau_{i, j} \}  ) ]^2 \nonumber \\
 &\leq&  \int^{\infty }_{ - \infty } \cdots \int^{\infty }_{ - \infty }
 ( \prod_{\langle i, j \rangle } d \tau_{i, j} ) \,
 x^2 ( \{ \tau_{i, j} \}  ) \cdot
 \int^{\infty }_{ - \infty } \cdots \int^{\infty }_{ - \infty }
 ( \prod_{\langle i, j \rangle } d \tau_{i, j} ) \,
 y^2 (\{ \tau_{i, j} \}  ) \, 
 \label{eq:Cauchy-SchwarzI}
\end{eqnarray}
 for example. When variables $x  ( \{ \tau_{i, j} \}  ) $ and  $y  ( \{ \tau_{i, j} \}  ) $ are
\begin{equation}
 x  ( \{ \tau_{i, j} \}  ) = \sqrt{ \frac{ 1}{(2 \pi J^2_1)^{\frac{N_B }{2 }} } \,
  e^{ - \sum_{\langle i, j \rangle }
  \frac{ \tau^2_{i, j} }{ 2 J^2_1 }
  -  \frac{ N_B ( K_b J_1 )^2 }{2}
  +  K_a
  \sum_{\langle i, j \rangle } \tau_{i, j} } }
 \, 
 \label{eq:GaJeft4}
\end{equation}
and
\begin{equation}
 y  ( \{ \tau_{i, j} \}  ) = \sqrt{ \frac{ 1}{(2 \pi J^2_1)^{\frac{N_B }{2 }} } \,
  e^{ - \sum_{\langle i, j \rangle }
  \frac{ \tau^2_{i, j} }{ 2 J^2_1 }
  -  \frac{ N_B ( K_b J_1 )^2 }{2}
  +  K_a
  \sum_{\langle i, j \rangle } \tau_{i, j} }  }
 \frac{Z (K_b )}{Z (K_a )   } 
 \, ,
 \label{eq:GaJeft5}
\end{equation}
 by applying the Cauchy-Schwarz inequality and the gauge transformation, we obtain
\begin{equation}
 \exp \biggl[ \frac{ N_B ( K^2_b - K^2_a ) J^2_1  }{2}  \biggr] \leq 
  [ \overline{e^{- \beta W}} ]_{K_b}
 \, .
 \label{eq:GaJeft6}
\end{equation}
 The left-hand side of Eq.~(\ref{eq:GaJeft6}) is a lower bound
 of the exponentiated work $[  \overline{e^{- \beta W}} ]_{K_b}$.
 The  lower bound of the exponentiated work $[  \overline{e^{- \beta W}} ]_{K_b}$
 does not depend on any lattice shape,
 and instead depends on the number of nearest-neighbor pairs in the whole system, $N_B$.
 This relation is rigorous, is useful for estimating the results by approximation
 approaches, and is useful for fixing bugs in computer programs
 for investigating this model.

 For the Sherrington-Kirkpatrick model, we obtain
\begin{equation}
 \exp \biggl[ \frac{(N - 1) (K^2_b - K^2_a ) J^2_1 }{4} \biggr] 
 \leq  
  [  \overline{e^{- \beta W}} ]_{K_b}
 \, .
 \label{eq:GaJeft7}
\end{equation}

 The nonequilibrium process for the exponentiated work $[ \overline{e^{- \beta W}} ]_{K_a}$
  is limited to the under left side of the Nishimori line in the phase diagram
 shown as the arrow `A' in Fig.~\ref{fig:phase-diagramGa}.
 Therefore, Eqs.~(\ref{eq:GaJeft2}) and (\ref{eq:GaJeft3})
 are useful for investigating the spin glass phase.
 The nonequilibrium process for the exponentiated work $[ \overline{e^{- \beta W}} ]_{K_b}$
  is limited to the upper right side of the Nishimori line in the phase diagram
 shown as the arrow `B' in Fig.~\ref{fig:phase-diagramGa}.
 Therefore, the left-hand sides of Ineqs.~(\ref{eq:GaJeft6}) and (\ref{eq:GaJeft7})
 are useful as lower bounds
 in investigating the ferromagnetic phase and the multicritical point
 for the paramagnetic phase, the ferromagnetic phase and the spin glass phase.

By using Eqs.~(\ref{eq:GaJeft2}) and (\ref{eq:GaJeft6}),
 we obtain
\begin{equation}
 [ \overline{e^{- \beta W}} ]_{K_a}  \leq
 [ \overline{e^{- \beta W}} ]_{K_b} \, .
 \label{eq:GaJeft8}
\end{equation}
 The relation (\ref{eq:GaJeft8})
 does not depend on any lattice shape,
 and also holds for the Sherrington-Kirkpatrick model.
 This relation is a rigorous relation between the exponentiated work
 which have different quenched random configurations.
 This relation is useful for understanding the exact exponentiated work
 in the phase diagram.
 This relation exactly shows that
 the exponentiated work of the nonequilibrium process shown
 as the arrow `B' in Fig.~\ref{fig:phase-diagramGa} is greater than or equal to
 that of the nonequilibrium process shown
 as the arrow `A' in Fig.~\ref{fig:phase-diagramGa}.

\section{Summary} \label{sec:5}

Exact results for the Jarzynski equality were derived
 for Ising spin glass models.
 As the Ising spin glass models, the $\pm J$ model and 
 the Gaussian model were investigated.
 For the  $\pm J$ model and the Gaussian model,
 we derived exact lower bounds of the exponentiated work
 for investigating the ferromagnetic phases and the multicritical points,
 and derived rigorous relations between the exponentiated work
 which have different quenched random configurations.
 For the Gaussian model,
 we derived the exact exponentiated work
 for investigating the spin glass phase.
 Exact results for the infinite-range models were also obtained.
 The present results were obtained by using a gauge symmetry.
 The present results are related to points on the Nishimori lines.
 The present results do not depend on any lattice shape, and
 a part of the present results instead depends on
 the number of nearest-neighbor pairs in the whole system.

The present results are useful
 for understanding the exact exponentiated work
 in the phase diagrams for the Ising spin glass models.
 In addition, the present results are
 useful for estimating the results by
 approximation approaches, and are useful for fixing bugs
 in computer programs for investigating the Ising spin glass models.

\end{document}